\documentclass[aps,pre,twocolumn,superscriptaddress,showpacs,showkeys,amsmath,amssymb]{revtex4}

\usepackage{epsfig,amsmath,amssymb,color}
\usepackage[T1]{fontenc}
\usepackage[latin9]{inputenc}
\usepackage{epstopdf}

\begin{document}

\title{Low-temperature thermodynamics of the two-leg ladder Ising model with trimer rungs: A mystery explained}

\author{Taras Hutak}
\affiliation{Institute for Condensed Matter Physics,
          National Academy of Sciences of Ukraine,
          Svientsitskii Street 1, 79011 L'viv, Ukraine}

\author{Taras Krokhmalskii}
\affiliation{Institute for Condensed Matter Physics,
          National Academy of Sciences of Ukraine,
          Svientsitskii Street 1, 79011 L'viv, Ukraine}

\author{Onofre Rojas}
\affiliation{Departamento de Fisica, 
          Universidade Federal de Lavras, 
          CP 3037, 37200-000, Lavras-MG, Brazil}

\author{Sergio Martins de Souza}
\affiliation{Departamento de Fisica, 
          Universidade Federal de Lavras, 
          CP 3037, 37200-000, Lavras-MG, Brazil}
          
\author{Oleg Derzhko}
\affiliation{Institute for Condensed Matter Physics,
          National Academy of Sciences of Ukraine,
          Svientsitskii Street 1, 79011 L'viv, Ukraine}

\date{\today}

\begin{abstract}
Recently, a surprising low-temperature behavior has been revealed in a two-leg ladder Ising model with trimer rungs
(Weiguo Yin, arXiv:2006.08921).
Motivated by these findings, we study this model from another perspective
and demonstrate that the reported observations are related to a critical phenomenon in the standard Ising chain.
We also discuss a related curiosity, namely, the emergence of a power-law behavior characterized by quasicritical exponents. 
\end{abstract}

\pacs{05.70.Fh, 75.10.-b, 75.10.Jm, 75.10.Pq}

\keywords{Ising model, two-leg ladder, thermodynamics, correlations}

\maketitle

\begin{figure}
\begin{center}
\includegraphics[clip=true,width=0.975\columnwidth]{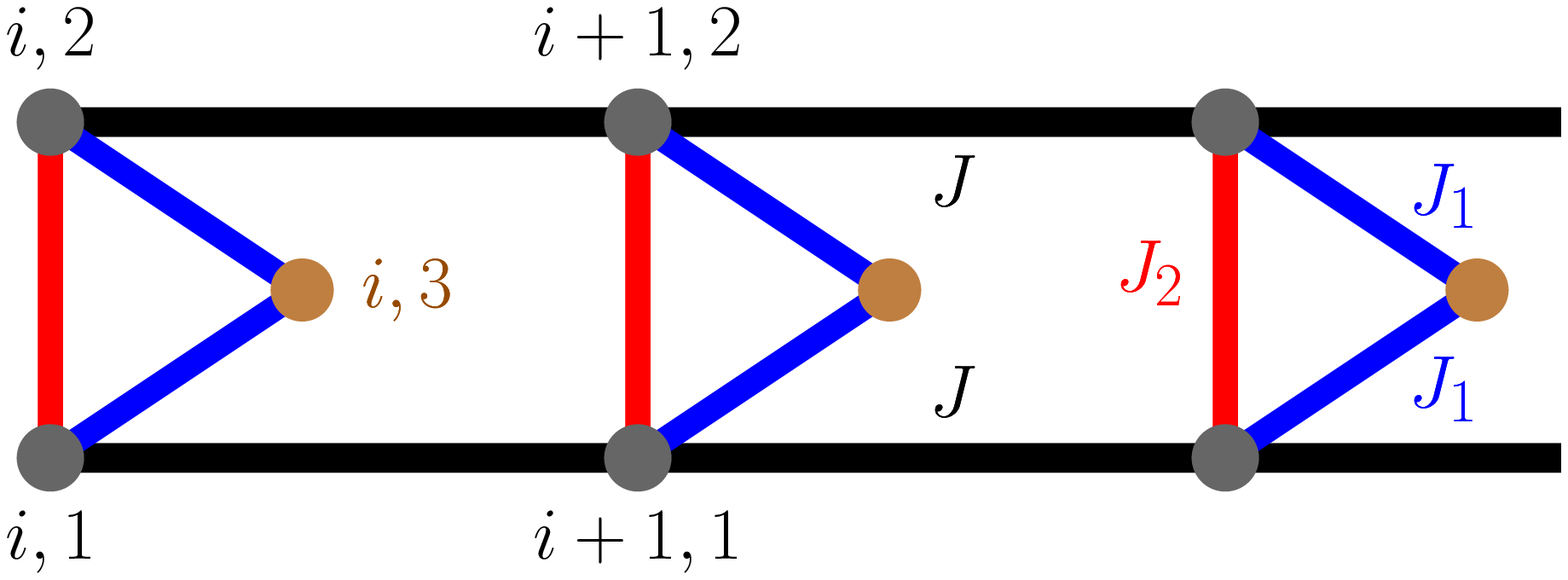}\\
\vspace{0mm}
\includegraphics[clip=true,width=0.975\columnwidth]{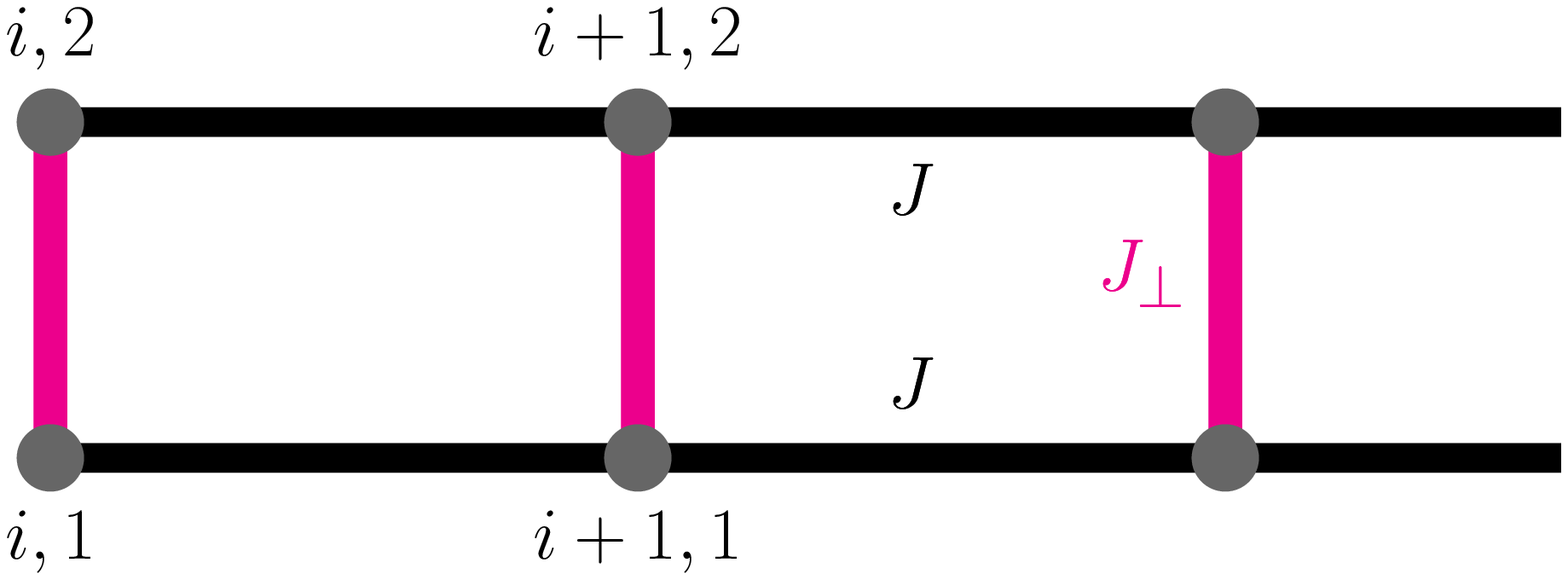}
\caption{(Top) The initial two-leg ladder Ising model with trimer rungs, see Eq.~\eqref{01}.
(Bottom) The effective two-leg (rail-road) ladder Ising model, see Eq.~\eqref{02}.
We denoted the bonds as follows: $J$ (black), $J_1$ (blue), $J_2$ (red), and $J_\perp$ (magenta).}
\label{figure01}
\end{center}
\end{figure}

Recently, Weiguo Yin considered several two-leg-ladder Ising models with short-range interactions 
which exhibit intriguing behavior at low temperatures which strongly resembles a finite-temperature phase transition \cite{Yin2020a,Yin2020b}.
A simplest model is defined by the Hamiltonian
\begin{eqnarray}
\label{01}
H=-\sum_{i=1}^{N}
\left[J\left(\sigma_{i,1}\sigma_{i+1,1}+\sigma_{i,2}\sigma_{i+1,2}\right)
\right.
\nonumber\\
\left.
+J_1\left(\sigma_{i,1}\sigma_{i,3}+\sigma_{i,2}\sigma_{i,3}\right)
+J_2\sigma_{i,1}\sigma_{i,2}
\right]
\end{eqnarray}
which is associated with the lattice shown in Fig.~\ref{figure01}, top.
Here 
$\sigma=\pm 1$, $N$ is the number of rungs, 
the positive/negative sign of the Ising exchange couplings corresponds to the ferromagnetic/antiferromagnetic interaction,   
and 
the exchange-coupling scheme is illustrated in Fig.~\ref{figure01}, top.
(Our $J_1$ and $J_2$ correspond to $J^{\prime}$ and $J^{\prime\prime}$ of Ref. \cite{Yin2020a}, respectively;
we have changed the notations 
since in what follows we use the prime notation to denote the first and second derivatives with respect to temperature.)
The model (\ref{01}) is exactly solvable by the transfer-matrix method \cite{Baxter1982,Yin2020a,Yin2020b}.
Weiguo Yin found that 
for $J_2<0$ and small positive values of the parameter $\alpha=(\vert J_1\vert-\vert J_2\vert)/\vert J\vert$
the exact specific heat exhibits a sharp peak at certain (low but finite) temperature, 
the exact entropy shows a waterfall behavior at this temperature etc \cite{Yin2020a}.  
He referred to such a phenomenon 
as marginal phase transition (MPT) \cite{Yin2020a,Yin2020b} or practically perfect phase transition (PPPT) \cite{Yin2020b}
and provided an extensive mathematical analysis to explain this phenomenon.
It is worthwhile noting 
that similar peculiar low-temperature thermodynamics was observed for other one-dimensional Ising-like models with short-range interactions
\cite{Galisova2015,Strecka2016,Torrico2016,Rojas2016,Souza2018,Carvalho2018,Rojas2018,Carvalho2019,Rojas2019,Rojas2020,Strecka2020a,Strecka2020b}.

In what follows,
we will explain the astonishing low-temperature behavior of the model (\ref{01}) using the approach suggested in Ref.~\cite{Krokhmalskii2019}.
Our present study is not restricted to model (\ref{01}),
but has a broader significance illustrating that the low-temperature peculiarities 
reported in Refs.~\cite{Yin2020a,Yin2020b,
Galisova2015,Strecka2016,Torrico2016,Rojas2016,Souza2018,Carvalho2018,Rojas2018,Carvalho2019,Rojas2019,Rojas2020,Strecka2020a,Strecka2020b}
all are related to the criticality of the standard Ising chain at zero temperature \cite{Baxter1982}.

The first step is 
to trace out in the partition function of the initial model (\ref{01}) the spins $\sigma_{i,3}$, $i=1,\ldots,N$ at the middle of the rung,
see Fig.~\ref{figure01}, top,
and 
to arrive at the effective model 
-- the standard two-leg rail-road Ising ladder 
with the interaction along the legs $-J$ and the temperature-dependent interaction that couples two legs $-J_\perp$,
see Fig.~\ref{figure01}, bottom
(decoration-iteration transformations \cite{Syozi1951}).
The Hamiltonian of the effective model reads:
\begin{eqnarray}
\label{02}
H=-NT\ln C 
\nonumber\\
-\sum_{i=1}^{N}
\left[J\left(\sigma_{i,1}\sigma_{i+1,1}+\sigma_{i,2}\sigma_{i+1,2}\right)+J_{\perp}\sigma_{i,1}\sigma_{i,2}\right],
\nonumber\\
C=2\sqrt{\cosh\frac{2J_1}{T}},
\;
J_\perp=J_\perp(T)=J_2+\frac{T}{2}\ln\cosh\frac{2J_1}{T}.
\end{eqnarray}
This representation was noticed in Ref.~\cite{Yin2020a} but not used for further analysis.

The thermodynamics of the model (\ref{02}) can be found by the transfer-matrix method \cite{Baxter1982}.
The Helmholtz free energy (per rung) $f$ reads:
\begin{eqnarray}
\label{03}
f=-T\ln C-T\ln\lambda_1,
\nonumber\\
\lambda_1
=
2\left(\!\cosh\frac{J_\perp}{T}\cosh\frac{2J}{T} \!+\! \sqrt{1+\sinh^2\frac{J_\perp}{T}\cosh^2\frac{2J}{T}}\!\right)\!.
\end{eqnarray}
The peculiarity of the effective ladder model (\ref{02}) stems from a dependence of the interleg coupling $-J_\perp$ on the temperature $T$.
Therefore, the internal energy $e$, the entropy $s$, and the specific heat $c$ are given by the formulas:
\begin{eqnarray}
\label{04}
e=-T^2\frac{\partial}{\partial T}\frac{f}{T}-T\frac{\partial f}{\partial J_\perp} J_\perp^{\prime}
=e^{(1)}+e^{(2)},
\nonumber\\
s=-\frac{\partial f}{\partial T} - \frac{\partial f}{\partial J_\perp} J_\perp^{\prime}
=s^{(1)}+s^{(2)},
\nonumber\\
c=-T\frac{\partial^2 f}{\partial T^2} 
\!-\! 2T\frac{\partial^2 f}{\partial T\partial J_\perp} J_\perp^{\prime}
\!-\! T \frac{\partial^2 f}{\partial J_\perp^2}\left(J_\perp^{\prime}\right)^2
\!-\! T \frac{\partial f}{\partial J_\perp} J_\perp^{\prime\prime}
\nonumber\\
=c^{(1)}+c^{(2)}+c^{(3)}+c^{(4)},
\end{eqnarray}
where according to Eq.~(\ref{03})
\begin{eqnarray}
\label{05}
\frac{\partial f}{\partial J_\perp}
=-\frac{\sinh\frac{J_\perp}{T}\cosh\frac{2J}{T}}{\sqrt{1+\sinh^2\frac{J_\perp}{T}\cosh^2\frac{2J}{T}}},
\nonumber\\
\frac{\partial^2 f}{\partial T\partial J_\perp}
=
\frac{J_\perp\cosh\frac{J_\perp}{T}\cosh\frac{2J}{T}+2J\sinh\frac{J_\perp}{T}\sinh\frac{2J}{T}}{T^2\left(1+\sinh^2\frac{J_\perp}{T}\cosh^2\frac{2J}{T}\right)^{\frac{3}{2}}},
\nonumber\\\
\frac{\partial^2 f}{\partial J_\perp^2}
=
-\frac{\cosh\frac{J_\perp}{T}\cosh\frac{2J}{T}}{T\left(1+\sinh^2\frac{J_\perp}{T}\cosh^2\frac{2J}{T}\right)^{\frac{3}{2}}},
\end{eqnarray}
and $J_\perp^{\prime}=\partial J_\perp/\partial T$, $J_\perp^{\prime\prime}=\partial^2 J_\perp/\partial T^2$.
It is worthy noting that 
$\partial f/\partial J_\perp=-\sum_{i=1}^N\langle \sigma_{i,1}\sigma_{i,2}\rangle /N \equiv -C_{12}(0)$,
where $C_{12}(0)$ is the on-rung correlation between two outer spins on the legs, see Ref.~\cite{Yin2020a}.
Moreover, for the correlation length $\xi$ we have
\begin{eqnarray}
\label{06}
\frac{1}{\xi}
=\ln
\frac{\cosh\frac{J_\perp}{T}\cosh\frac{2J}{T} + \sqrt{1+\sinh^2\frac{J_\perp}{T}\cosh^2\frac{2J}{T}}}
{\cosh\frac{J_\perp}{T}\cosh\frac{2J}{T} - \sqrt{1+\sinh^2\frac{J_\perp}{T}\cosh^2\frac{2J}{T}}}.
\end{eqnarray}

\begin{figure}
\begin{center}
\includegraphics[clip=true,width=0.975\columnwidth]{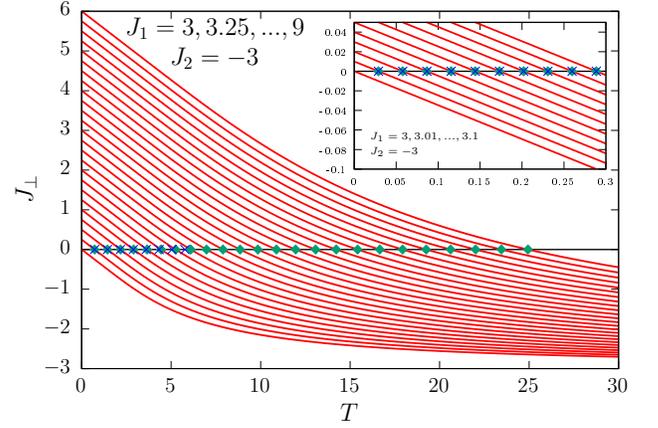}
\caption{Dependence of $J_\perp$ on $T$ as it follows from Eq.~(\ref{02})
for the set of parameters 
$J_1=3,3.01,\ldots,3.1$ (from bottom to top in the inset)
or 
$J_1=3,3.25,\ldots,9$ (from bottom to top in the main panel)
and $J_2=-3$ in Eq.~(\ref{01}).
Green diamonds correspond to $T_p$.
Blue crosses correspond to the approximate analytical formula 
$T_p= 2(\vert J_1\vert - \vert J_2\vert)/\ln 2$, 
which is valid if 
$J_2=-\vert J_2\vert<0$ 
and 
$(\vert J_1\vert -\vert J_2 \vert)/\vert J_1\vert\to +0$.}
\label{figure02}
\end{center}
\end{figure}

Equations (\ref{02}) -- (\ref{06}) explain the enigmatic low-temperature behavior of the initial model (\ref{01}).
Assume that the effective interleg interaction in Eq.~(\ref{02}) 
is ferromagnetic for $0\le T< T_p$, vanishes at $T=T_p$, and becomes antiferromagnetic for $T_p<T$,
see Fig.~\ref{figure02} for examples.
The equation 
\begin{eqnarray}
\label{07}
J_\perp(T_p)=0
\end{eqnarray}
has a simple analytical solution for $T_p>0$
if $J_2=-\vert J_2\vert<0$ and $(\vert J_1\vert -\vert J_2 \vert)/\vert J_1\vert\to +0$.
Namely, 
$T_p=2(\vert J_1\vert - \vert J_2\vert)/\ln 2\approx 2.885(\vert J_1\vert - \vert J_2\vert)$.
This is the MPT temperature reported in Ref.~\cite{Yin2020a}.
At $T=T_p$ the legs decouple, see Eq.~\eqref{07}, and the correlation length (\ref{06}) becomes
$1/\xi(T_p)=\ln([\cosh(2J/T_p) + 1]/[\cosh(2J/T_p) - 1])$,
cf. the correlation length for the standard Ising chain \cite{Baxter1982}.
It is not surprising that $\xi(T_p)$ is extremely large if $T_p/\vert J\vert$ is small enough,
i.e., 
the correlation length for the standard Ising chain at the temperature $T_p$ which is low in the scale of $\vert J\vert$ is obviously large
[in this limit $\xi(T_p)\propto\exp(2\vert J\vert/T_p)$].
Hence, 
if
\begin{eqnarray}
\label{08}
\frac{T_p}{\vert J\vert}\ll 1 
\end{eqnarray}
holds, 
one observes the traces of the Ising-chain criticality in the behavior of the initial model (\ref{01})
which were interpreted as very unusual and puzzling phenomena.
For above mentioned specific case 
$J_2=-\vert J_2\vert<0$ and $(\vert J_1\vert -\vert J_2 \vert)/\vert J_1\vert\to +0$,
the inequality (\ref{08}) becomes $(\vert J_1\vert - \vert J_2\vert)/\vert J\vert=\alpha\ll 1$,
i.e., corresponds to the strong frustration regime of Ref.~\cite{Yin2020a}.
Substituting $J_\perp(T_p)=0$ into Eq.~(\ref{05})
we obtain:
$\partial f/\partial J_\perp=\partial^2 f/\partial T\partial J_\perp=0$,
and
$\partial^2 f/\partial J_\perp^2=-[\cosh(2J/T_p)]/T_p$.
If Eq.~(\ref{08}) holds, 
this results in a large value of the specific heat $c(T_p)$, 
see the third term $c^{(3)}$ in the formula for $c$ in Eq.~(\ref{04}) which is $\propto\exp(2\vert J\vert/T_p)(J_\perp^\prime)^2$.

Let us consider $T$ in the vicinity of $T_p$ 
when the inequality $\sinh^2(J_\perp/T)\cosh^2(2J/T)\gg 1$ holds.
Then, 
according to Eq.~(\ref{05}),
$\partial f/\partial J_\perp\approx {\rm {sgn}}(J_\perp)$
whereas
$\partial^2 f/\partial T\partial J_\perp\propto {\rm {sgn}}(J_\perp)/\vert J_\perp\vert^2$
and
$\partial^2 f/\partial J_\perp^2 \propto 1/\vert J_\perp\vert^3$.
Moreover,
according to Eq.~(\ref{06}),
$1/\xi\propto \vert J_\perp\vert$.
Bearing in mind that $ J_\perp \propto T-T_p$ around $T_p$ (see Fig.~\ref{figure02}),
one immediately concludes that while approaching $T_p$ 
(i) $e$ and $s$ show a jump, 
see the second terms $e^{(2)}$ and $s^{(2)}$ in the formulas for $e$ and $s$ in Eq.~(\ref{04}), 
and
(ii)
$c$ and $\xi$ show power-law dependences: $c\propto \vert T-T_p\vert^{-3}$ and $\xi\propto \vert T-T_p\vert^{-1}$,
see the third term $c^{(3)}$ in the formula for $c$ in Eq.~(\ref{04}) and Eq.~(\ref{06}).
Such quasicritical exponents 
$\alpha=\alpha^\prime=3$ and $\nu=\nu^\prime=1$
were observed for similar one-dimensional Ising-like models in Ref.~\cite{Rojas2019}.
It is worthy noting 
that the exactly calculated specific heat $c$ [Eq.~(\ref{04})]
satisfies the sum rule:
$\int_0^\infty{\rm{d}}T c/T=s(\infty)-s(0)$,
where $s(\infty)=3\ln 2\approx 2.079$ and $s(0)=0$ are the entropies at infinite and zero temperatures, respectively.
Hence,
the higher the peak of $c$ around $T_p$ is, a narrower it should be.
Recall also that precisely at $T_p$ both $c$ and $\xi$ are finite,
see above.

\begin{figure}
\begin{center}
\includegraphics[clip=true,width=0.975\columnwidth]{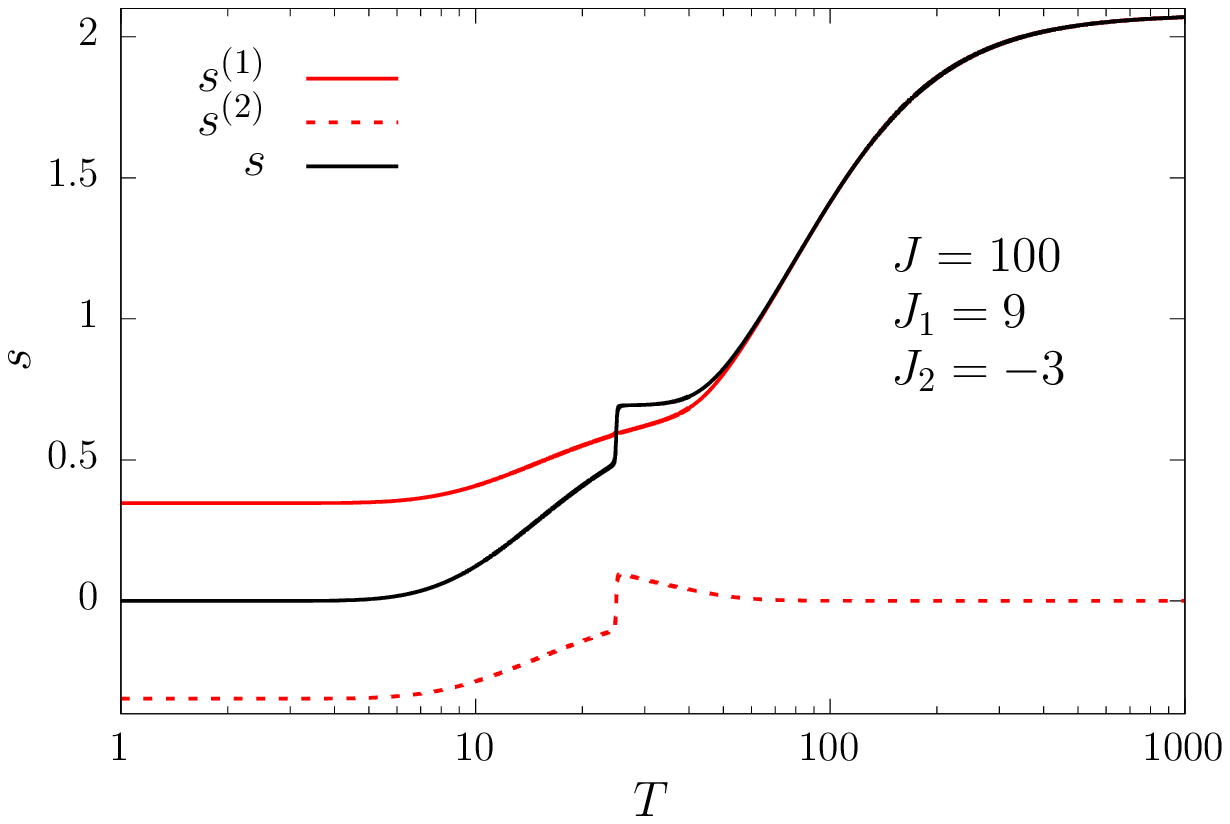}\\
\vspace{0mm}
\includegraphics[clip=true,width=0.975\columnwidth]{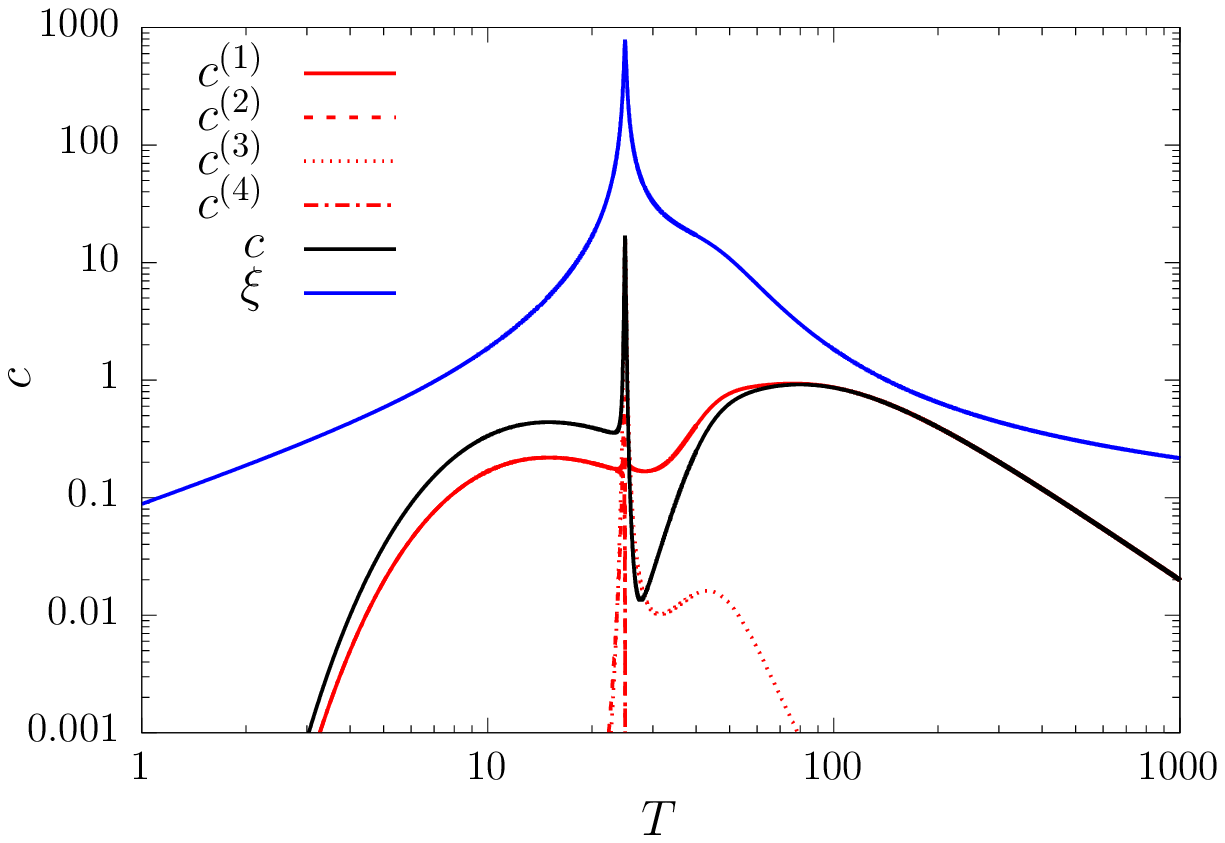}\\
\vspace{0mm}
\includegraphics[clip=true,width=0.975\columnwidth]{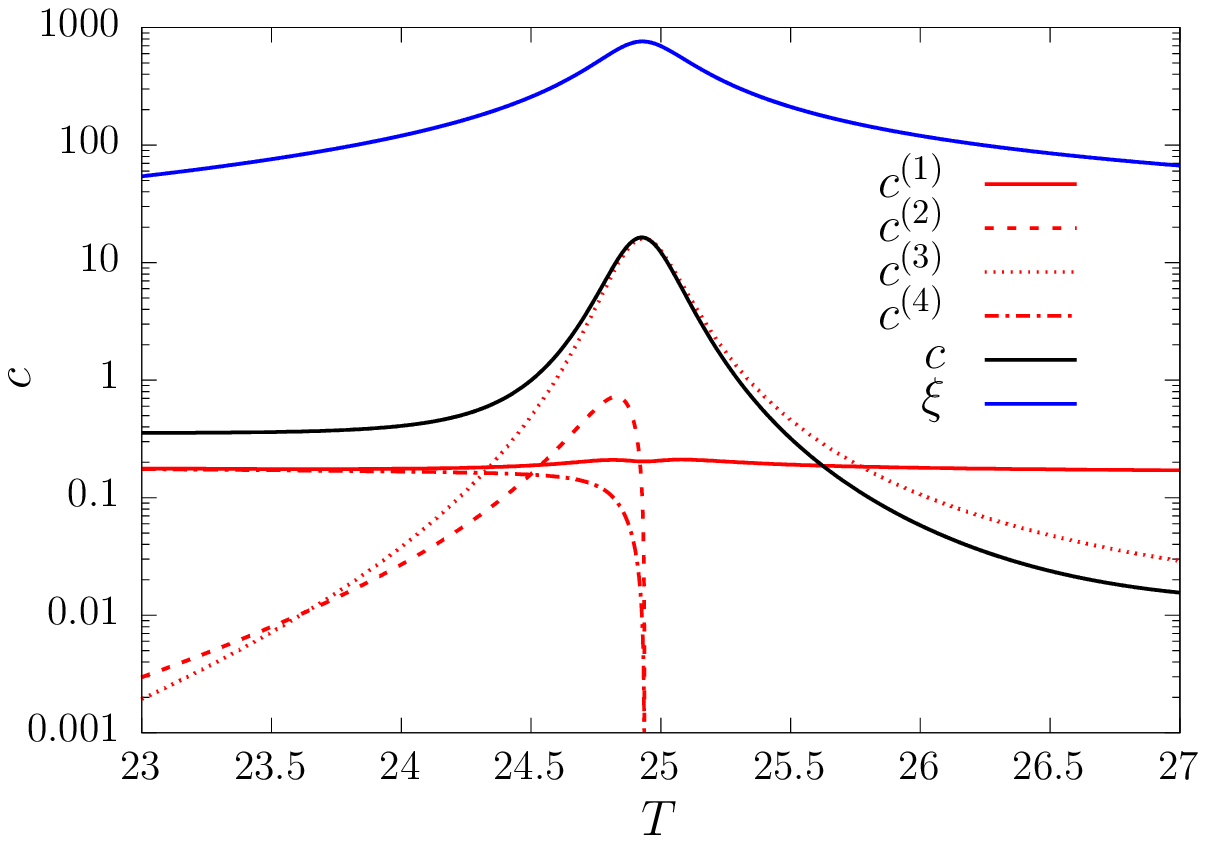}
\caption{Low-temperature properties of the two-leg ladder Ising model with trimer rungs,
see Eq.~(\ref{01}) and Fig.~\ref{figure01}, top, 
for the set of parameters $J_1=9$, $J_2=-3$, and $J=100$.
(Top) Entropy versus temperature in the 1--1000 temperature range. 
(Middle) Specific heat and correlation length versus temperature in the 1--1000 temperature range.
(Bottom) Specific heat and correlation length versus temperature in the vicinity of $T_p\approx 24.937$.
The contribution of different terms to the entropy and the specific heat, see Eq.~(\ref{04}), is shown explicitly.}
\label{figure03}
\end{center}
\end{figure}

\begin{figure}
\begin{center}
\includegraphics[clip=true,width=0.975\columnwidth]{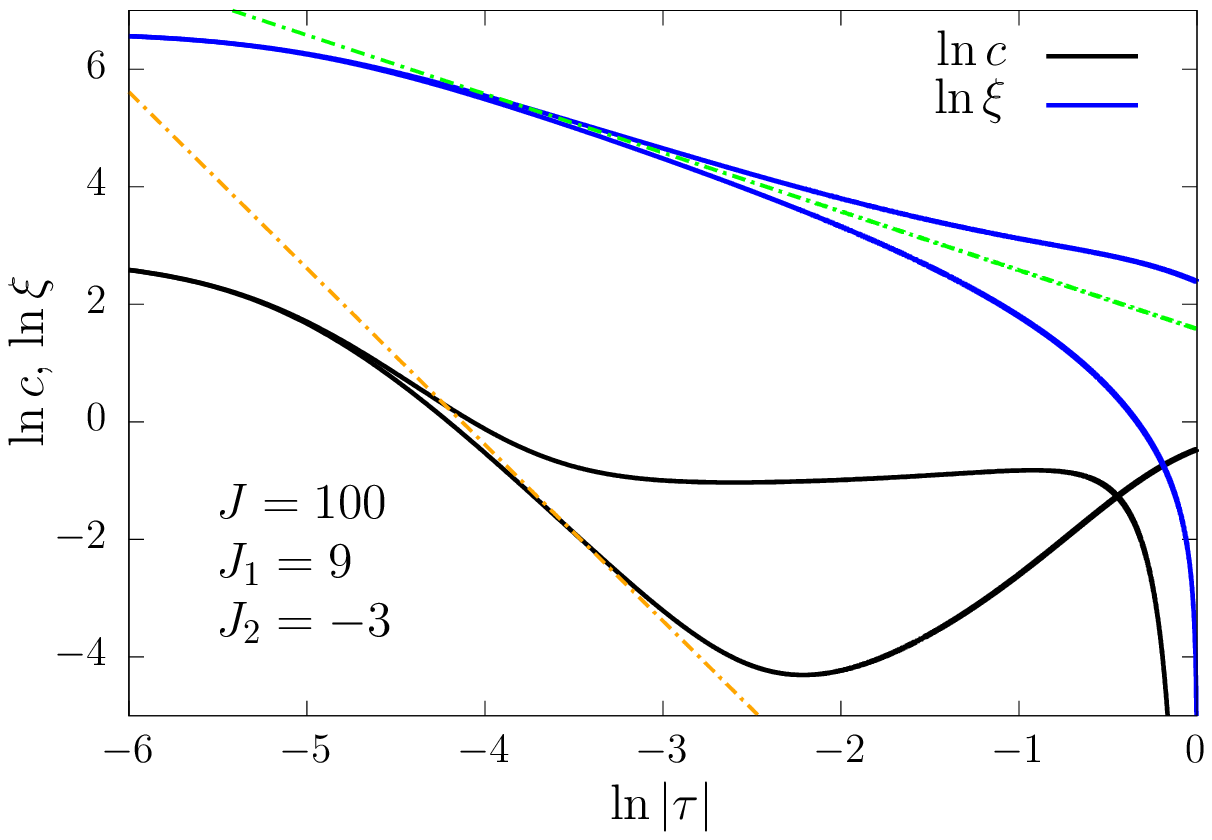}\\
\vspace{0mm}
\includegraphics[clip=true,width=0.975\columnwidth]{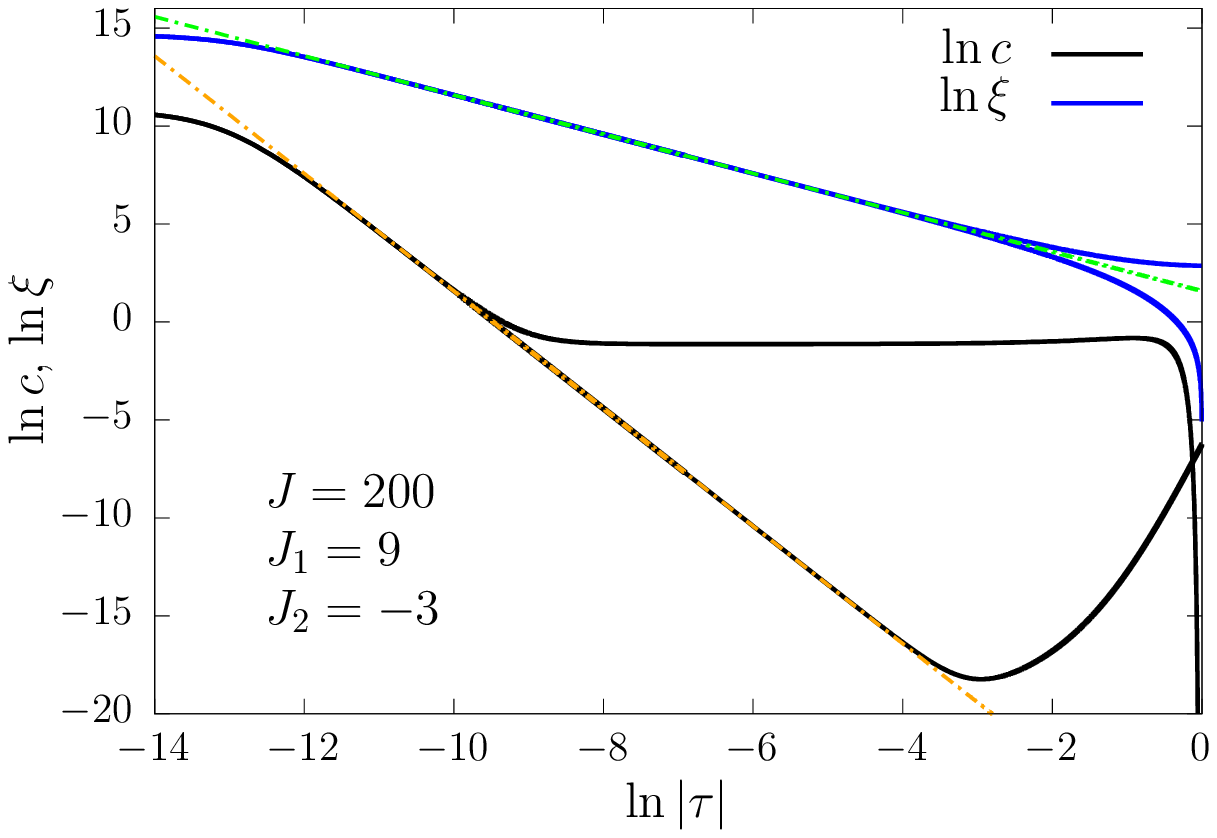}
\caption{Quasicritical exponents for the specific heat and the correlation length
as evidenced by the dependences 
$\ln c$ (ascending and descending parts of the peak, black curves) 
and 
$\ln \xi$ (ascending and descending parts of the peak, blue curves) 
on $\ln\vert\tau\vert$, $\tau=(T-T^*)/T^*$, 
$T^*$ stands for the temperature at which $c(T)$ or $\xi(T)$ has the peak.
The set of parameters is as follows:
$J_1=9$, $J_2=-3$ ($T_p\approx 24.937$)
and
(top) $J=100$ 
[then $T_c\approx 24.927$, $c(T_c)\approx 16.4$ and $T_\xi\approx 24.929$, $\xi(T_\xi)\approx 7.61\cdot 10^2$]
or
(bottom) $J=200$ 
[then $T_c\approx 24.937$, $c(T_c)\approx 4.88\cdot 10^4$ and $T_\xi\approx 24.937$, $\xi(T_\xi)\approx 2.31 \cdot 10^6$].
Dashed lines correspond to power-law dependences with the exponent 3 (gold) and the exponent 1 (green).}
\label{figure04}
\end{center}
\end{figure}

To illustrate the discussion,
we consider a specific example conveniently setting $J_1=9$, $J_2=-3$, and $J=100$. 
In this case 
the frustration parameter $\alpha=0.06$ 
whereas $T_p\approx 24.937$ yielding $T_p/\vert J\vert \approx 0.249$ in Eq.~(\ref{08})
[since $(\vert J_1\vert-\vert J_2\vert)/\vert J_1\vert\approx 0.667$, the analytical formula underestimates the value of $T_p$].
$J_\perp$ as a function of $T$ for this set can be seen in the main panel of Fig.~\ref{figure02}.
The entropy $s$ [Eq.~(\ref{04})] exhibits a jump at $T_p$ due to the term $s^{(2)}=-(\partial f/\partial J_\perp)J^{\prime}_\perp$,
$s(T_p+0)-s(T_p-0)\approx 0.2$,
see the black curve in Fig.~\ref{figure03}, top.
The specific heat $c$ [Eq.~(\ref{04})] exhibits a peak around $T_p$ due to the term $c^{(3)}=-T(\partial^2 f/\partial J_\perp^2)(J_\perp^\prime)^2$,
$c(T_p)\approx 16.3$,
see the black curves in Fig.~\ref{figure03}, middle and bottom.
The correlation length $\xi$ [Eq.~(\ref{06})] also exhibits a peak around $T_p$,
$\xi(T_p)\approx 760$,
see the blue curves in Fig.~\ref{figure03}, middle and bottom.
Since $\vert J_\perp^{\prime}(T_p)\vert<1$, we have $c(T_p)<\xi(T_p)$.
Finally,
Fig.~\ref{figure04} demonstrates quasicritical behavior:
There is a finite range of temperatures in the vicinity of $T_p$
on which a clear power-law behavior develops 
with $\alpha=\alpha^{\prime}=3$ for the specific heat (black curves) and $\nu=\nu^{\prime}=1$ for the correlation length (blue curves).
Moreover,
the smaller is $T_p/\vert J\vert$ in the left-hand side of the inequality in Eq.~(\ref{08}), 
the larger the region of quasicriticality is,
cf. the top panel ($T_p/\vert J\vert\approx0.249$) and the bottom panel ($T_p/\vert J\vert\approx0.125$) in Fig.~\ref{figure04}. 
Recall that the quasicritical behavior fails for temperatures in the immediate vicinity of $T_p$ where both $c$ and $\xi$ are finite,
see the values of $c$ and $\xi$ at the smallest values of $\tau$ in Fig.~\ref{figure04}.

Let us summarize our findings.
First of all, we have to emphasize the following.
The model at hand is exactly solvable one 
(as well as other models discussed in Refs.~\cite{Yin2020a,Yin2020b,Galisova2015,Strecka2016,Torrico2016,Rojas2016,Souza2018,Carvalho2018,
Rojas2018,Carvalho2019,Rojas2019,Rojas2020,Strecka2020a,Strecka2020b}) 
and thus there is no much room for speculations about the solutions:
The performed transformations are rigorous and the derived results are exact.
However, the final results are somewhat astonishing:
For example, 
the specific heat per cell $c$ may exhibit an extra low-temperature peak of extremely large height,
the correlation length $\xi$ at this temperature is unexpectedly large etc,
see Figs.~\ref{figure03} and \ref{figure04}.
Naturally, these features of the rigorously found quantities call for explanations.

As a rule, the authors sought for explanations at the level of the initial models.
In contrast, 
in Ref.~\cite{Krokhmalskii2019} we suggest to examine, instead of several diverse initial models, the effective model
which shows up after summing over redundant degrees of freedom.
The effective model 
that has emerged after this rigorous decoration-iteration transformation \cite{Syozi1951}
is the Ising chain with temperature-dependent parameters. 
In the present study,
applying this way of thinking to a typical representative considered in Refs.~\cite{Yin2020a,Yin2020b}
-- the two-leg ladder Ising model with trimer rungs --
we have arrived at the two-leg Ising ladder with temperature-dependent rung couplings $J_\perp=J_\perp(T)$,
Eq.~\eqref{02}.
Again,
the analysis of the properties of the obtained effective two-leg Ising ladder with temperature-dependent parameters 
explains the low-temperature thermodynamics of a whole class of the initial models 
which can be cast into that effective model after eliminating redundant spins.
 
Although we do not deny the usefulness of studies on the level of the initial model,
however, we see several advantages of the study based on the effective model.
First, we clearly see a universality of the phenomenon:
Many initial models collapse to one effective model 
and, 
after all, 
the low-temperature peculiarities 
of two-leg-ladder Ising models \cite{Yin2020a,Yin2020b}
as well as 
of other one-dimensional Ising-like models \cite{Krokhmalskii2019}
are related to the criticality of the standard Ising chain at zero temperature \cite{Baxter1982}.
From our consideration it looks that frustration is not vitally necessary 
(see also Ref.~\cite{Rojas2018}); 
the only demand is to have a suitable $J_\perp(T)$.
The characteristic temperature  $T_p$ is determined by the condition when the effective model reduces to the standard Ising-chain model,
see Eq.~(\ref{07}).
Temperature-dependent parameters of the effective model result in more complex formulas for thermodynamic quantities, see Eq.~(\ref{04}).
If $T_p$ is small in the scale of the effective Ising-chain model,
Eq.~\eqref{08},
the large values of the specific heat or the correlation length are evident. 
Moreover, 
as it follows from Eqs.~(\ref{04}), (\ref{05}), (\ref{06}) and Eq.~(\ref{02}),
the temperature dependences around $T_p$ 
(but not in the immediate vicinity of $T_p$)
follow power laws with quasicritical exponents $\alpha=\alpha^{\prime}=3$ and $\nu=\nu^{\prime}=1$.

Finally,
our work paves the path for the higher-dimensional cases.
For example,
one can consider a two-dimensional rectangular effective Ising model 
with the horizontal couplings $J$ and the vertical couplings $J_\perp$ given in Eq.~(\ref{02}).
Obviously, if condition (\ref{07}) holds, 
the two-dimensional system becomes a set of noninteracting Ising chains and it exhibits the behavior discussed above.

The authors gratefully acknowledge helpful correspondence with Weiguo Yin.
T.~Hutak was supported by the fellowship of the National Academy of Sciences of Ukraine for young scientists.
O.~Rojas and S.~M.~de~Souza thank CNPq (Brazil) and FAPEMIG (Brazil) for partial financial support.


\begin{thebibliography}{99}

\bibitem{Yin2020a}
Weiguo Yin,
{\it Frustration-driven unconventional phase transitions at finite temperature in a one-dimensional ladder Ising model},
arXiv:2006.08921v2.

\bibitem{Yin2020b}
Weiguo Yin,
{\it Finding and classifying an infinite number of cases of the practically perfect phase transition in an Ising model in one dimension},
arXiv:2006.15087v1.

\bibitem{Baxter1982}
R.~J.~Baxter, 
{\it Exactly Solved Models in Statistical Mechanics}
(Academic Press, 1982).

\bibitem{Galisova2015}
L.~G\'{a}lisov\'{a} and J.~Stre\v{c}ka,
Phys. Rev. E {\bf 91}, 022134 (2015).

\bibitem{Strecka2016}
J.~Stre\v{c}ka, R.~C.~Al\'{e}cio, M.~L.~Lyra, and O.~Rojas, 
J. Magn. Magn. Mater. {\bf 409}, 124 (2016).

\bibitem{Torrico2016}
J.~Torrico, M.~Rojas, S.~M.~de~Souza, and O.~Rojas, 
Phys. Lett. A {\bf 380}, 3655 (2016).

\bibitem{Rojas2016}
O.~Rojas, J.~Stre\v{c}ka, and S.~M.~de~Souza, 
Solid State Communications {\bf 246}, 68 (2016).

\bibitem{Souza2018}
S.~M.~de~Souza and O.~Rojas,
Solid State Communications {\bf 269}, 131 (2018).

\bibitem{Carvalho2018}
I.~M.~Carvalho, J.~Torrico, S.~M.~de~Souza, M.~Rojas, and O.~Rojas,
J. Magn. Magn. Mater. {\bf 465}, 323 (2018).

\bibitem{Rojas2018}
O.~Rojas, 
{\it A conjecture on the relationship between critical residual entropy and finite temperature pseudo-transitions of one-dimensional models},
arXiv:1810.07817v2.

\bibitem{Carvalho2019}
I.~M.~Carvalho, J.~Torrico, S.~M.~de~Souza, O.~Rojas, and O.~Derzhko,
Annals of Physics {\bf 402}, 45 (2019).

\bibitem{Rojas2019}
O.~Rojas, J.~Stre\v{c}ka, M.~L.~Lyra, and S.~M.~de~Souza,
Phys. Rev. E {\bf 99}, 042117 (2019).

\bibitem{Rojas2020}
O.~Rojas, J.~Stre\v{c}ka, O.~Derzhko, and S.~M.~de~Souza,
J. Phys.: Condens. Matter {\bf 32}, 035804 (2020).

\bibitem{Strecka2020a}
J.~Stre\v{c}ka,
Acta Physica Polonica A {\bf 137}, 610 (2020).

\bibitem{Strecka2020b}
J.~Stre\v{c}ka,
{\it Pseudo-critical behavior of spin-1/2 Ising diamond and tetrahedral chains},
arXiv:2002.06942v1.

\bibitem{Krokhmalskii2019}
T.~Krokhmalskii, T.~Hutak, O.~Rojas, S.~M.~de~Souza, and O.~Derzhko,
{\it Towards low-temperature peculiarities of thermodynamic quantities for decorated spin chains},
arXiv:1908.06419v1.

\bibitem{Syozi1951}
I.~Syozi, 
Prog. Theor. Phys. {\bf 6}, 341 (1951); 
M.~Fisher, 
Phys. Rev. {\bf 113}, 969 (1959); 
O.~Rojas, J.~S.~Valverde, and S.~M.~de~Souza, 
Physica A {\bf 388}, 1419 (2009); 
J.~Stre\v{c}ka, 
Phys. Lett. A {\bf 374}, 3718 (2010); 
O.~Rojas and S.~M.~de~Souza, 
J. Phys. A: Math. Theor. {\bf 44}, 245001 (2011).

\end{thebibliography}
\end{document}